\documentclass[]{aa}
\usepackage{amsmath}
\usepackage{txfonts} 
\usepackage{graphicx}
\usepackage{subfigure}
\usepackage{color}
\usepackage{bm}
\usepackage{natbib}
\usepackage{url}
\usepackage[center, labelfont=bf, justification=justified]{caption}
\usepackage{ragged2e}

\usepackage{tabularx}
\usepackage{graphicx}
\usepackage{txfonts}
\usepackage{longtable}
\usepackage{hhline}
\usepackage{arydshln}
\usepackage{multirow}
\usepackage{multicol}
\usepackage{lscape}
\usepackage{array}
\usepackage{rotating}
\usepackage{float}

\newcommand{\msun}{$M_\odot$}

\usepackage{gensymb}

\newcommand{\bra}{19:37:00.9}
\newcommand{\bdec}{+07:34:09.6}
\newlength{\pointwidth}
\newcommand{\AnaelleProj}{2016.1.01552.S}
\newcommand{\hfs}{\textit{HfS}}

\newcommand{\co}{$^{12}$CO}
\newcommand{\coseven}{C$^{17}$O}
\newcommand{\coeight}{C$^{18}$O}

\newcommand{\hidrogen}{H$_{2}$}
\newcommand{\cth}{C$_{2}$H}

\begin{document}

\title{Structured velocity field in the inner envelope of B335: ALMA observations of rare CO isotopologues}

  \author{Victoria Cabedo  
  	       \inst{1,3},
	      Ana\"elle Maury
  	       \inst{1,2},
  	      Josep M. Girart
  	       \inst{3,4},
  	      Marco Padovani
  	        \inst{5}
  	       }

\institute{
Astrophysics department, CEA/DRF/IRFU/DAp, Universit\'{e} Paris Saclay, UMR AIM, F-91191 Gif-sur-Yvette, France \\
\email{victoria.cabedo@cea.fr} 
\and
Harvard-Smithsonian Center for Astrophysics, 60 Garden street, Cambridge, MA 02138, USA 
\and
Institut de Ci\`encies de l'Espai (ICE), CSIC,
Can Magrans s/n, Cerdanyola del Vall\`es, 08193 Catalonia, Spain
\and
Institut d'Estudis Espacials de Catalunya (IEEC), 08034 Barcelona, Catalonia, Spain
\and
INAF-Osservatorio Astrofisico di Arcetri, Largo E. Fermi 5, 50125 Firenze, Italy
}

\abstract
{Studying Class 0 objects is very important, as it allows to characterize dynamical processes at the onset of the star formation process, and to determine the physical mechanisms responsible for the outcome of the collapse. Observations of dense gas tracers allow the characterization of key kinematics of the gas directly involved in the star-formation process, such as infall, outflow or rotation.}
{This work aims at investigating the molecular line velocity profiles of the Class 0 protostellar object B335 and attempts to put constraints on the infall motions happening in the circumstellar gas of the object.}
{Observations of \coseven\ (1-0), \coeight\ (1-0) and \co\ (2-1) transitions are presented and the spectral profiles are analyzed at envelope radii between 100 and 860 au.}
{\coseven\ emission presents a double peaked line profile distributed in a complex velocity field. Both peaks present an offset of 0.2 to 1 km s$^{-1}$ from the systemic velocity of the source in the probed area. The optical depth of the \coseven\ emission has been estimated and found to be less than 1, suggesting that the two velocity peaks trace two distinct velocity components of the gas in the inner envelope.}
{After discarding possible motions that could produce the complex velocity pattern, such as rotation and outflow, it is concluded that infall is producing the velocity field. Because inside-out symmetric collapse cannot explain those observed profiles, it is suggested that those are produced by non-isotropic accretion from the envelope into the central source along the outflow cavity walls.}


\authorrunning{V. Cabedo et al.}
\titlerunning{Structured velocity field in the envelope of B335}
\maketitle


\section{Introduction} 
    
    Low-mass stars are known to form in dense molecular gas clouds. Class 0 objects represent the first stage of the star formation process, when most of the mass is still contained in the envelope surrounding the protostar (\citealt{Andre1993}, \citealt{Andre1995}). Models of protostellar collapse (\citealt{Shu1987}) suggest that it is during this phase that the circumstellar gas is transported to the central object thanks to accretion processes. During this stage, angular momentum needs to be removed from the envelope and stored in the central object or dissipated through viscous processes to allow the formation of the star. Moreover, the accretion mode, the rate at which it happens, and the duration of possible accretion episodes during this phase will determine the final stellar mass (\citealt{Andre1995}, \citealt{Basu2004}, \citealt{Bate2005}, \citealt{Myers2012}). Therefore, studying this phase is crucial, as it allows to understand which are the kinematics and dynamics of the gas at the onset of collapse, and to determine how those affect the outcome of the star formation process. 
    
    The gas making most of the protostellar envelopes is typically probed using molecular gas line profiles, which trace the gas kinematics in the dense envelope and are used to measure gas motions such as rotation or infall. Observations of the molecular line emission from embedded protostars have been suggested to trace widespread infall signatures in the inner $\sim$ 2000 au of some protostellar envelopes (\citealt{Zhou1993}, \citealt{Rawlings1996}, \citealt{DiFrancesco2001}, \citealt{Mottram2013}). Most of those studies rely on the detection and interpretation of the infall spectral signature known as blue asymmetry or inverse P-Cygni profile. In the center of a dense protostellar core, where gas densities become larger as the collapse proceeds ($>$ 10$^{5}$ cm$^{-3}$, \citealt{Andre1995}), the large optical depth of some molecular line emission produce a self-absorbed line profile, with the dip centered on the emission at systemic velocity, where most of the circumstellar gas emits. Because the core is collapsing under the effect of gravity, line emission from dense gas tracers will present the typical blue-asymmetry line profile. Most works focused on modeling these line profiles to put constraints on protostellar infall models, rely on the assumption of a symmetrical cloud collapsing, which produces such double-peaked profile on optically thick emission, to extract key information such as central protostellar mass, infall velocities and mass accretion rates (\citealt{Zhou1993}, \citealt{DiFrancesco2001}, \citealt{Evans2005}, \citealt{Evans2015}). However, blue-asymmetries in line profiles are not unique to infall motions. Complex gas kinematics (\citealt{Maureira2017}), such as asymmetric collapse (\citealt{Tokuda2014}), accretion streamers (\citealt{Pineda2020}, \citealt{Segura-Cox2020}) or outflow-entrained gas, can produce separated velocity components on the same line-of-sight which are observed as double-peaked line profiles not caused by optical thickness.
    
    The isolated Bok globule B335, which contains an embedded Class 0 protostar \citep{Keene1983} is located at a distance of 164.5 pc, \citep{Watson2020} and has been the prototypical object to test symmetrical collapse infall models, since blue-asymmetries were first detected in molecular emission of the source at core scales \citep{Zhou1993,Choi1995,Evans2005}. Double-peaked line profiles have also been observed with interferometric observations of the molecular emission from the inner envelope (\citealt{Chandler1993}; \citealt{Saito1999}; \citealt{Yen2010}; \citealt{Kurono2013}; \citealt{Evans2015}). Infall models in an optically thick line emission have attempted to compute infall mass rates (\citealt{Yen2010}, \citealt{Evans2015}, \citealt{Yen2015b}), obtaining values affected by large uncertainties that range from 10$^{-7}$ M$_\odot$yr$^{-1}$ to $\sim$3$\times$10$^{-6}$ M$_\odot$yr$^{-1}$ at radii of 100-2000 au, and infall velocities from 1.5 km s$^{-1}$ to $\approx$0.8 km s$^{-1}$ at radii of $\sim$ 100 au. New models based on continuum emission and using the revised distance of 164.5 pc  have determined an infall rate from the envelope to the disk of 6.2$\times$10$^{-6}$ M$_\odot$yr$^{-1}$ (Evans et al., in prep.). For the estimated age of 4$\times$10$^{4}$ yr, this implies a total mass at the center (star $+$ disk) of 0.26 M$_\odot$. The physical cause of these double peaked line profiles has been questioned, however. For example \citet{Kurono2013} pointed out that despite expecting the H$^{13}$CO$^{+}$ emission should be optically thin, the inverse p-Cygni profile and the position-velocity diagram they observe can be reproduced with models of moderately optically thick infalling gas.
    B335 is associated with an east-west outflow, prominently detected in \co\, with an inclination of 10\degr\ on the plane of the sky and an opening angle of 45\degr\ (\citealt{Hirano1988}, \citealt{Hirano1992}, \citealt{Yen2010}). The eastern lobe is slightly oriented on the near side (\citealt{Stutz2008}) producing blueshifted emission on the eastern side and redshifted emission on the western side. While the core has been found to be slowly rotating at large scales ($>$ 2500 au) (\citealt{Frerking1987}, \citealt{Saito1999}, \citealt{Yen2010}, \citealt{Yen2011}), no clear rotation was found at smaller radii ($<$ 1000 au) and no kinematic signature of a disk was reported down to $\sim$ 10 au (\citealt{Yen2015b}, \citealt{Yen2018}). Recent observations of the hour-glass shaped magnetic field at small scales have suggested that B335 is an excellent candidate for magnetically regulated collapse (\citealt{Maury2018}). 
    
    In this work, observations of the molecular lines \coseven\ (1-0) and \coeight\ (1-0), which trace the dense circumstellar gas of the inner envelope of B335, are presented along with the \co\ (2-1) line emission, tracing the outflow cavity. The molecular line profiles are analyzed and interpreted, giving new constraints on the gas kinematics close to the protostar.

\section{Observations and data reduction}

    Observations of the Class 0 protostellar object B335 were carried out with the ALMA interferometer during the cycle 4 observation period from October 2016 to September 2017, as part of the project \AnaelleProj\ (PI. A. Maury). In all the work, it is assumed that the centroid position of B335 is at $\alpha = $ \bra\ and $\delta = $ \bdec\ in J2000 coordinates, corresponding to the peak of dust continuum obtained from high resolution maps (\citealt{Maury2018}). All lines were targeted using a combination of ALMA configurations: \coseven\ (1-0) and \coeight\ (1-0) were targeted using two configurations, C40-2 and C40-5, and \co\ (2-1) was targeted using C40-1 and C40-4. Technical details of the observations are shown in Table \ref{table:technicalInfo}.
    
    Preliminary analysis of the data was done with the product images delivered by ALMA to check if emission was detected and to check the shape of the line profiles. The \coseven\ emission was only detected in the most compact configuration (C40-2), therefore only this configuration has been used to produce \coseven\ and \coeight\ maps, while \co\ was detected in both configurations so a combination of the two data sets has been used to produce the maps. Calibration of the raw data was done using the standard script for cycle-4 ALMA data using the Common Astronomy Software Applications (CASA) version 5.6.1-8. Continuum emission was self-calibrated with CASA. Line emission was calibrated using the self-calibrated model derived from the continuum data when it was possible. Final images of the data were generated from the calibrated visibilities using the tCLEAN algorithm within CASA, using Briggs weighting with robust parameter set to 2 for \coseven\ and \coeight\, and 1 for \co. After imaging, \co\ maps were smoothed to reach the same angular  resolution as \coseven\ and \coeight. The resulting map characteristics are shown in Table  \ref{table:ImageChar}.

    \begin{table}[!ht]
        \centering
            \caption{Maps characteristics.}
        \begin{tabular}{c c c c}
            \hline
            \hline
            & & \\
            Transition & \co\ (2-1) & \coseven\ (1-0) & \coeight (1-0) \\
            Rest Frequency (GHz) & 230.53 & 112.359 & 109.782\\
            Beam major (arcsec) & 2.6 & 2.6 & 2.7 \\
            Beam minor (arcsec) & 1.9 & 1.9 & 2.0 \\
            PA (\degree) & 88 & 88 & 88 \\
            Spectral Res. (km~s$^{-1}$) & 0.20 & 0.20 & 0.20\\
            LRS* (arcsec) & 10.6 & 21.9 & 22.4 \\
            RMS (mJy/beam) & 2016.9 & 10.0  &  10.7 \\
            & & &  \\
            \hline
        \end{tabular}
        \justify
        \begin{list}{}{}
            \item * Largest recoverable scale computed as $\Theta_{LRS} = (0.6\lambda/B_{min})206265$ in arcsec, where $\lambda$ is the rest wavelength of the line, in m, and $B_{min}$ is the minimum baseline of the configuration, in m (\citealt{ALMAc4}).
        \end{list}
        \label{table:ImageChar}
    \end{table}

    \begin{figure}
        \begin{tabular}{m{8.5cm}}\includegraphics[width=9cm]{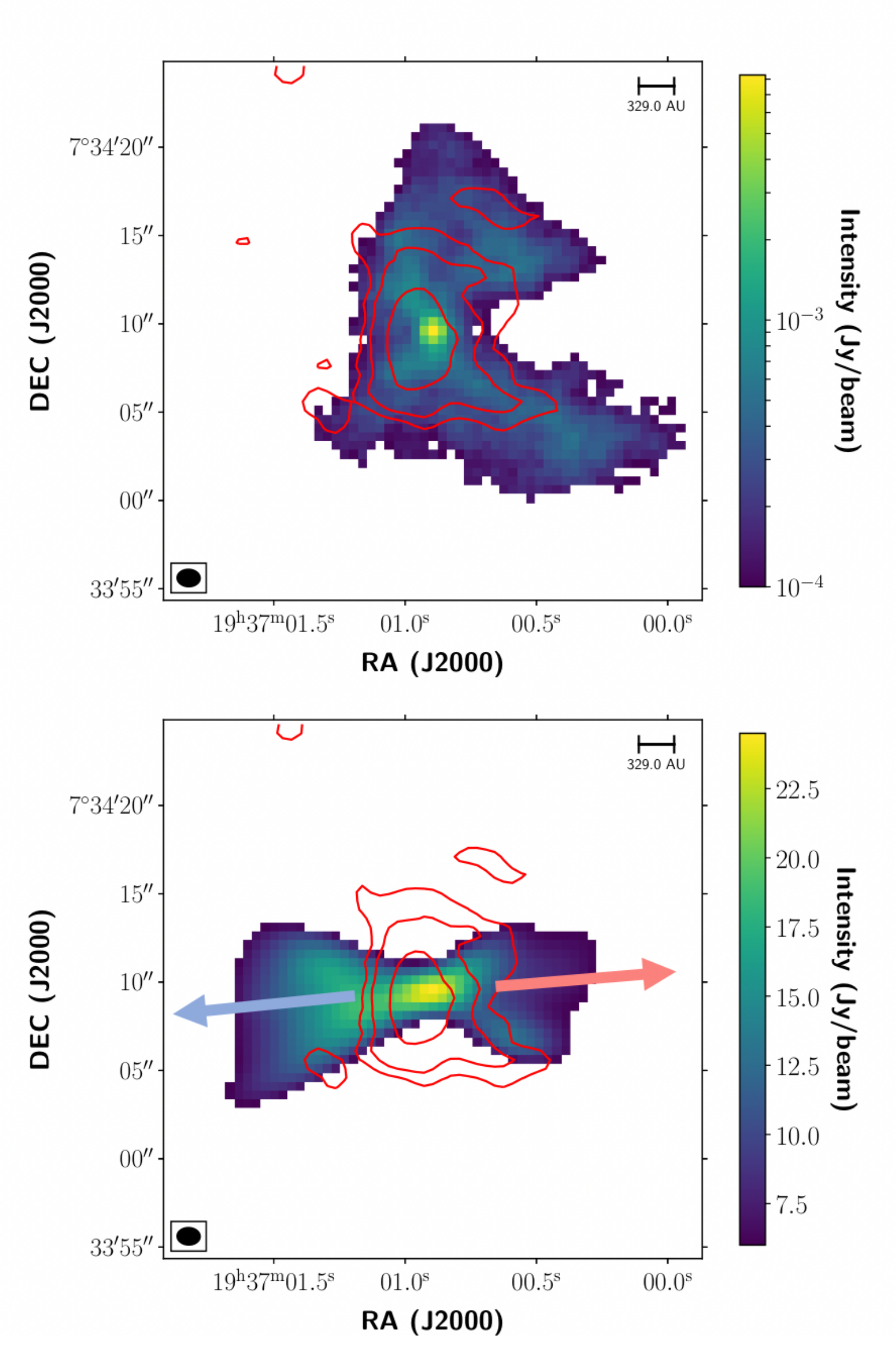} \\
        \end{tabular}
        \caption{Red contours showing moment 0 of \coseven\ emission, integrated over the velocity range 4.8-6.2 km~s$^{-1}$ and 7.6-9.4 km~s$^{-1}$. Contours show emission at $-$3, 3, 5, 10 and 30 $\sigma$, where $\sigma$ is 10.0 mJy/beam. Top: Intensity shows dust continuum emission map at 110 GHz for emission over 3$\sigma$, where $\sigma$ is 8.56$\times10^{-2}$ mJy/beam. Bottom: Intensity shows \co\ (2-1) moment 0 emission integrated over the velocity range 1.4-16.2 km~s$^{-1}$, for emission over 3$\sigma$, where $\sigma$ is 2016.9 mJy/beam. The two arrows show the direction of the E-W outflow.}
        \label{fig:C17O_intensity}
    \end{figure}

\section{Results and analysis}

    Figure \ref{fig:C17O_intensity} shows the moment 0 of \coseven\ (1-0) emission in red contours. The mean radius of the 3$\sigma$ emission is 860 au, indicating that it traces the dense envelope. Top image shows in intensity the dust continuum map observed at 110 GHz, showing emission over 3$\sigma$. The bottom image shows in intensity the moment 0 of \co\ (2-1) emission. The \co\ emission probes the outflowing gas, therefore it is confirmed that the \coseven\ emission is tracing the envelope and it is not affected by the outflow. 
    
\begin{figure*}
        \centering
        \includegraphics[width=\textwidth]{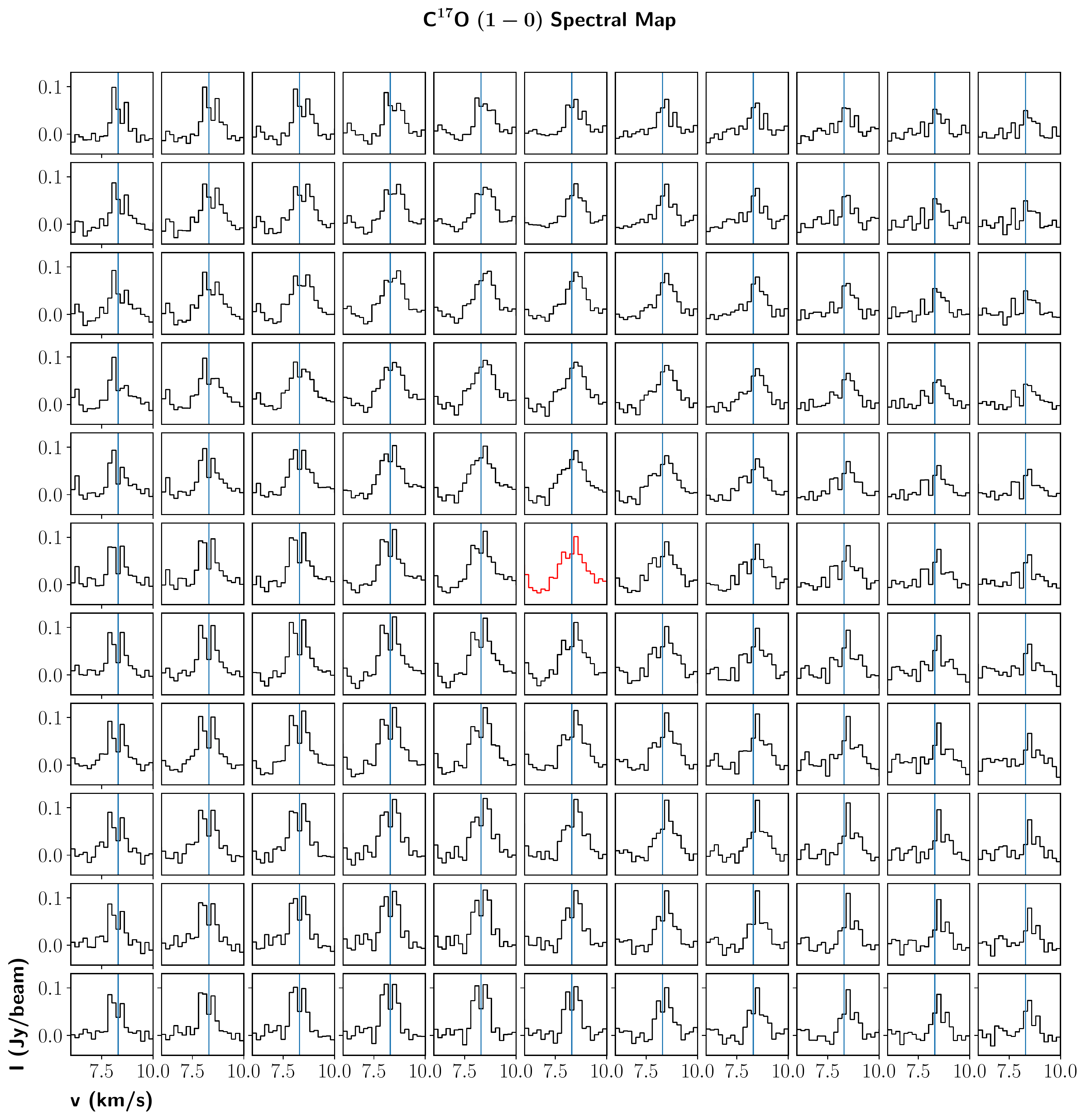} \\
        \caption{Spectral map of the \coseven\ (1-0) emission in the inner 900 au, centered on the dust continuum emission peak. The whole map is 5.5" $\times$ 5.5" and each pixel correspond to 0.5" ($\sim$ 82 au). For clarity, the spectral range only shows the main hyperfine component. The green spectrum refers to the peak of the continuum emission and the blue line indicates the systemic velocity (8.3 km~s$^{-1}$).}
        \label{fig:lineProf_c17o}
    \end{figure*}

\begin{figure*}
        \centering
        \includegraphics[width=0.55\textwidth, angle=90, trim={2.5cm 0 2cm 0}, clip]{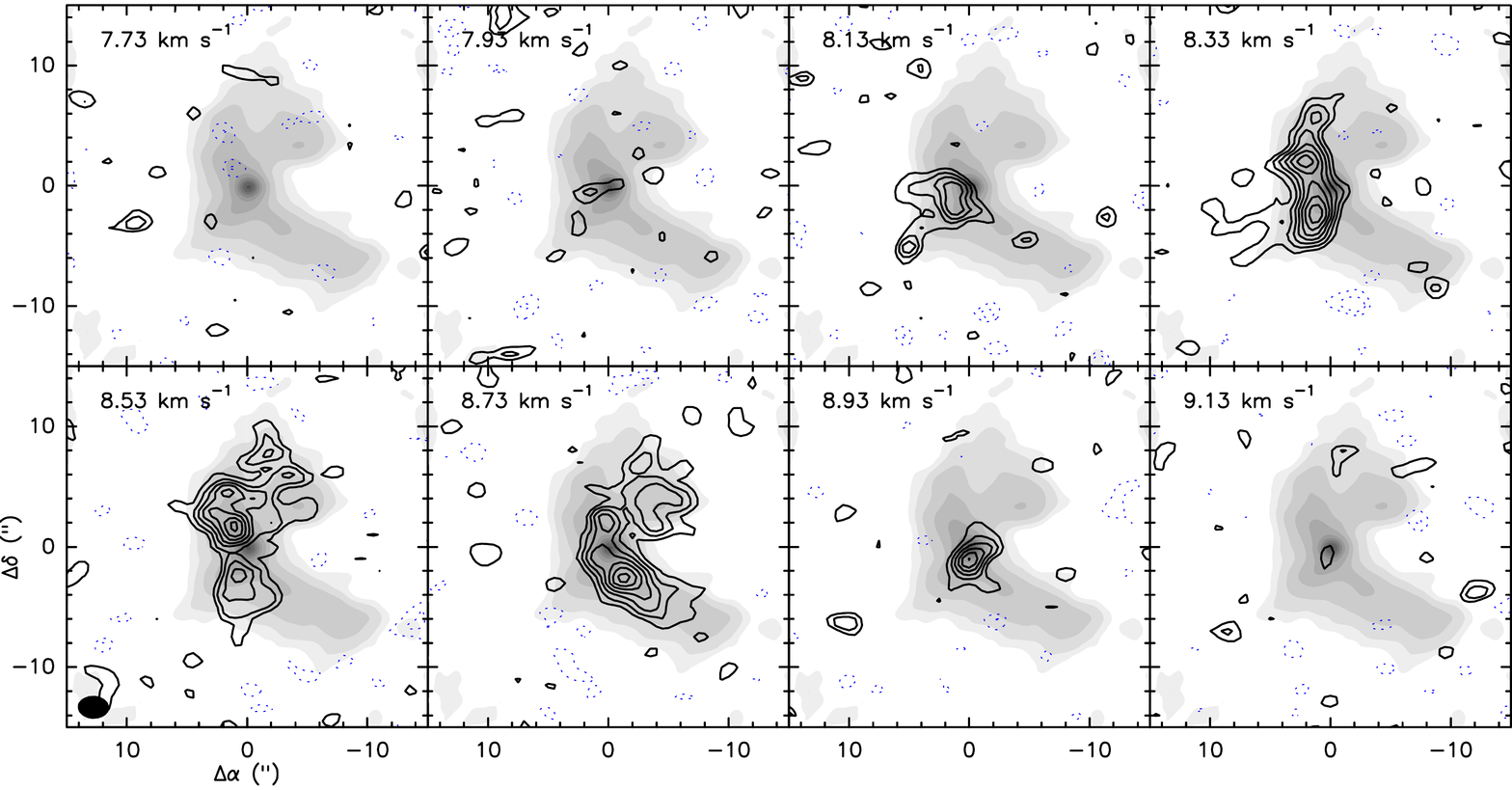} \\
        \caption{Contour channel maps of the emission of the  F=5/2--($-$5/2) hyperfine component of the \coseven\ J=1-0 transition, overlapped with the gray scale image of the 1.3 mm dust emission. Contours are $-4$, $-2$, and from 2 to 18 by steps of 2 times the rms noise of the channel maps, 10~mJy~Beam$^{-1}$. The synthesized beam is shown in the bottom left panel as a filled ellipse. The $v_{\rm LSR}$ channel velocities are shown in the top left part of the panels.}
        \label{fig:c17o_hfs_channelMaps}
    \end{figure*}

    In order to understand the dynamics of the gas that is being probed, the \coseven\ spectra were taken at every 0.5\arcsec\ pixel of the emission cube, producing a spectral map shown on Fig. \ref{fig:lineProf_c17o}. The line profile patterns show two distinct velocity components with a dip centered around the systemic velocity (8.3 km s$^{-1}$). Their respective intensities vary depending on the direction of the offset from the continuum peak, being the blue component more intense in the Eastern part of the core, while the red component is dominant in the Western part. This behavior is true for all the detected hyperfine components. A clear broadening of the line can be observed near the dust continuum emission peak, and in the North-East region part of the core. The former can be due to natural thermal broadening of the two velocity components as the temperature rises in the center of the object, while the latter might be the consequence of the overlapping of the two components due to other dynamical processes. The possibility of the dip being caused by interferometric filtering is discarded since the recovered emission size is of the order of the Largest Recoverable Scale (see Table \ref{table:ImageChar}). Moreover, because \coseven\ is a rare isotopologue, it is not expected to be abundant at largest scales and therefore no emission can be filtered at the systemic velocity. Figure \ref{fig:c17o_hfs_channelMaps} shows the velocity channel maps of the \coseven\ (1-0) emission, covering a velocity range from 7.7 to 9.1 km/s and only showing the main hyperfine component (Fig. \ref{fig:channel_maps_fullHFS} shows a range covering from 5.0 to 9.6 km/s, showing the two observed hyperfine components). It can be seen that the blue- and red-shifted emission are confined to the east and west regions, respectively, suggesting that the two components are probing gas with different dynamics.
    
    Three independent methods have been used to investigate the origins of the \coseven\ spectral profiles and if the two distinct peaks could be produced by an optically thick line emission. In the following sections, the analysis of these methods and the modeling of the velocity field in the B335 inner envelope are presented.

    \subsection{Line opacity estimation}
    
        The maximum opacity at the center of the source has been estimated from the \hidrogen\ column density and assuming a standard \coseven\ abundance using Eq. \ref{eq:opacity} (\citealt{Jansen1995}).
        
        \begin{equation}
            \tau_{0} = \frac{A_{10}^{C^{17}O} c^{3}}{8\pi \nu^{3}}\frac{N_{0}^{H_{2}}[C^{17}O]}{\Delta V}
            \label{eq:opacity}
        \end{equation}
        
        Where $A_{10}^{C^{17}O}$ = 6.695 $\times$ 10$^{-8}$ s$^{-1}$ is the Einstein coefficient for the \coseven\ J = 1--0 transition (\citealt{Hitran2017}, \citealt{Splatalogue2005}), $c$ is the speed of light, $\nu$ is the frequency of this transition,  $N_{0}^{H_{2}}$ = 3.1 $\times$ 10$^{22}$ cm$^{-2}$ is the peak column density of \hidrogen\ in B335 at a radius of 3600 au (\citealt{Launhardt2013}), [C$^{17}$O] = 5 $\times$ 10$^{-8}$ is the \coseven\ abundance relative to \hidrogen\ abundance (\citealt{Thomas2008}) and $\Delta V$ $\approx$ 1 kms$^{-1}$ is the average observed linewidth for the two components together. Using these values the obtained opacity is $\tau_{0}$ = 0.77, which corresponds to an opacity typical from an optically thin line.

     \subsection{Intensity ratio}

        Because of their similar mass and molecular structure, \coseven\ and \coeight\ should probe gas under similar physical conditions. The [\coseven]/[\coeight] isotope ratio does not appear to be affected by fractionation and, if emission from both \coseven\ and \coeight\ is associated to dense gas shielded from external ultraviolet radiation, selective photo-dissociation probably does not affect the relative abundances (\citealt{vanDishoeck1988}). Thus, the only difference in the emission from these two isotopes should result from opacity effects, because \coeight\ is a factor of 3.6-3.9 more abundant than \coseven\ (\citealt{Penzias1981}, \citealt{Jorgensen2002}). The ratio of integrated intensities is much less sensitive to linewidth effects than the ratio of peak intensity, therefore we use the latter to rule out any abundances effect on the \coseven\ emission.

        We produced beam-matching maps for \coseven\ and \coeight\ to compare the gas at similar scales in both isotopes. The obtained synthesized beams are given in Table \ref{table:ImageChar}. Intensities are integrated over the two velocity ranges of 4.8-6.2 km~s$^{-1}$ and 7.4-9.4 km~s$^{-1}$ for \coseven\, hence taking into account the two observed hyperfine components, and 7.4-9.4 km~s$^{-1}$ for \coeight. The fact that both lines emit in the same range of velocities and are spatially coincident indicates that they are probing the same reservoir of circumstellar gas. 
        
\begin{figure}
            \begin{tabular}{m{8.5cm}}\includegraphics[width=9cm]{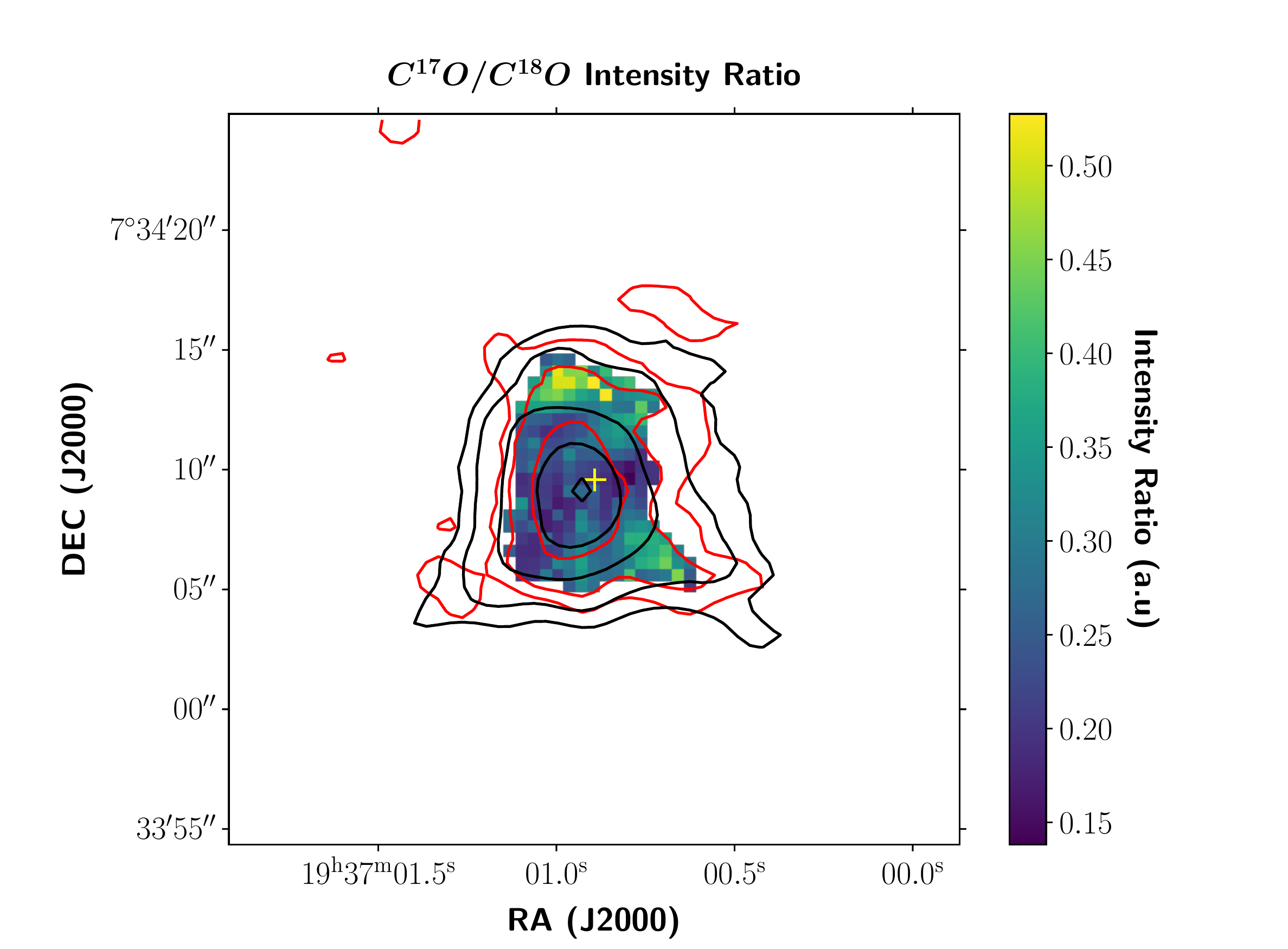} \\
            \end{tabular}
            \caption{\coseven\ to \coeight\ integrated intensity ratio. Red contours show integrated intensity for \coseven\ at $-$3, 3, 5, 10 and 30 $\sigma$, where $\sigma$ is 10.0 mJy/beam. Black contours show integrated intensity for \coeight\ at -3, 3, 5, 10 and 30$\sigma$, where $\sigma$ is 10.7 mJy/beam. The yellow cross indicates the centroid position of B335.}
            \label{fig:c17o_c18o_ratio}
        \end{figure}

        The integrated intensity ratio was computed as W$_{C^{17}O}$/W$_{C^{18}O}$, where W$_{i}$ is the integrated intensity of each \coseven\ and \coeight\ individual spectra. Figure \ref{fig:c17o_c18o_ratio} shows the obtained integrated intensity ratio map, with values ranging from 0.15 to 0.50 and a mean value centered at 0.28.  We note that the intensity ratio is quite homogeneous in most of the extension of the emission, but gets larger at the North and South-West regions. This is attributed to a line broadening of the \coseven\ in those regions when compared to the \coeight\ emission (see \coeight\ spectral map, Fig. \ref{fig:lineProf_c18o}). The origin of this broadening is unknown, but we attribute it to complex dynamical processes that might be taking place in these regions.
        
        The expected ratio if both transitions are optically thin is computed as [\coseven]/[\coeight] $\approx$ 0.25, where [\coseven] and [\coeight] are respectively the \coseven\ and \coeight\ abundance with respect to \hidrogen\ ([\coseven] = 5 $\times 10^{-8}$ and [\coeight] = 2 $\times 10^{-7}$, \citealt{Thomas2008}). Our observations are therefore in general agreement with the expected ratio if both transitions are optically thin. No increase of the intensity ratio is seen on the observations towards the center of the source, where opacity is expected to be higher, further confirming the later hypothesis. Note that this is also in agreement with the conclusions reached by the analysis of single-dish observations of the $\rm C^{18}O\ (2-1)$ and $\rm C^{17}O\ (2-1)$ ($\tau _{\rm C^{18}O\ (2-1)} \sim 0.8$, \citealt{Evans2005}).

    \subsection{Modeling of the molecular line profiles}
        
\begin{figure*}[!ht]
            \centering
            \includegraphics[width=17cm]{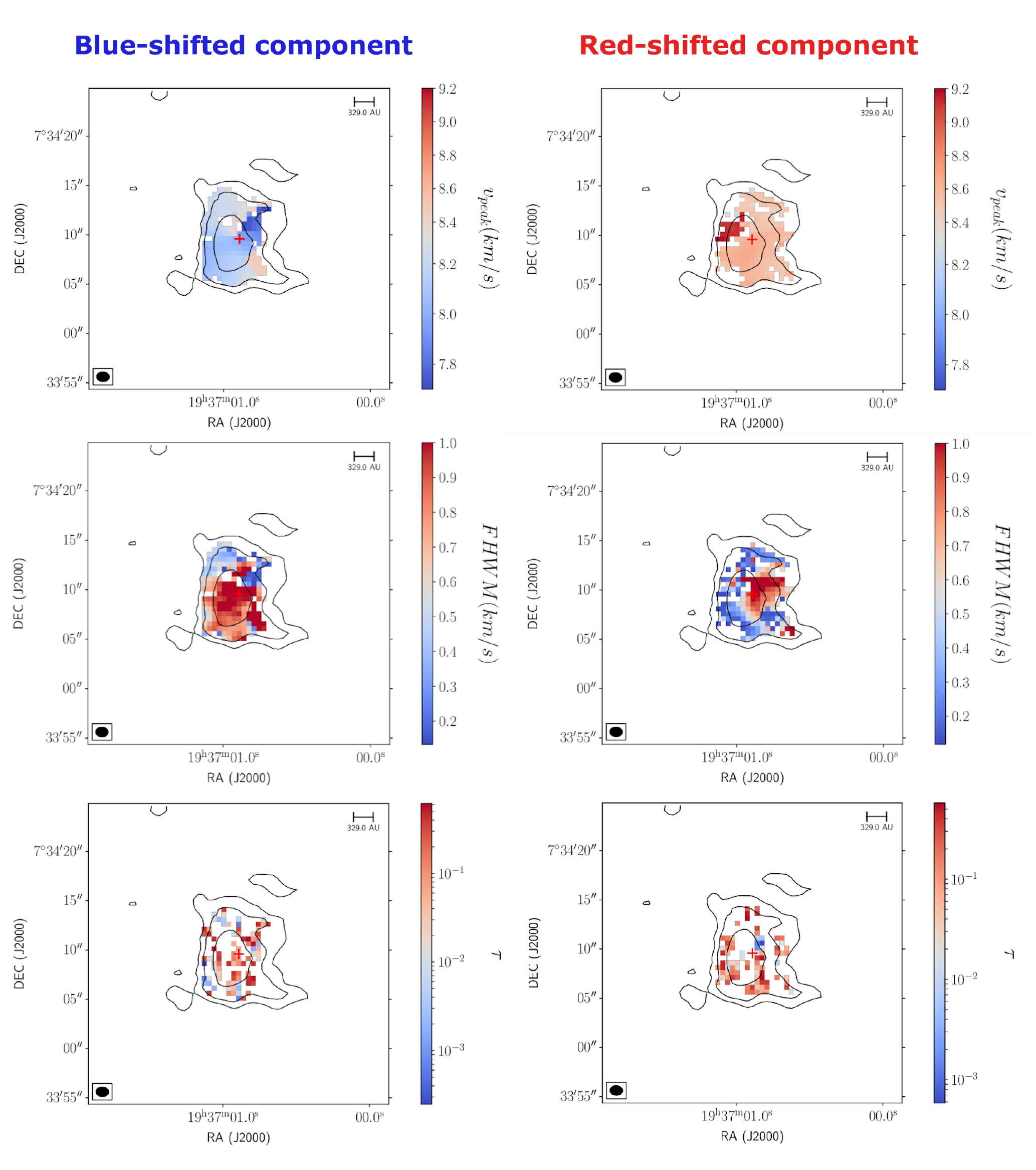} \\
            \caption{\coseven\ (1-0) maps obtained from modeling the line profiles with two velocity components. Overlaid contours show the integrated intensity at 5, 10, 20 and 30 $\sigma$, where $\sigma$ is 10.0 mJy/beam. Top: peak velocity; middle: linewidth and bottom: opacity. Each column shows one of the two velocity components. Red crosses indicate the centroid position of B335.}
            \label{fig:c17o_fullmaps}
        \end{figure*}

        The spectrum at each pixel has been modeled using a program that allows to fit the hyperfine structure of spectral lines with multiple velocity components (\hfs, \citealt{Estalella2017}). It also allows to compute the opacity of the line from the relative intensity of the different hyperfine components of the given transition. For every velocity component the general \hfs\ procedure fits simultaneously four independent parameters: the linewidth assumed to be the same for each hyperfine component; the main line central velocity; the main line peak intensity and the optical depth. The fitting procedure samples the space parameters to find the minimum value of the fit residual $\chi^{2}$. Because the \coseven\ (1-0) emission is expected to be optically thin, it was attempted to model the double-peaked profiles with two velocity components, one blue- and another red-shifted. Initial expected values were introduced and the program is allowed to proceed to fit emission with a minimum signal-to-noise ratio of 3$\sigma$. The peak velocity, the linewidth and the opacity maps obtained from the fitting are shown in Fig. \ref{fig:c17o_fullmaps}.

\begin{figure}
            \begin{tabular}{m{8.5cm}}\includegraphics[width=8.5cm]{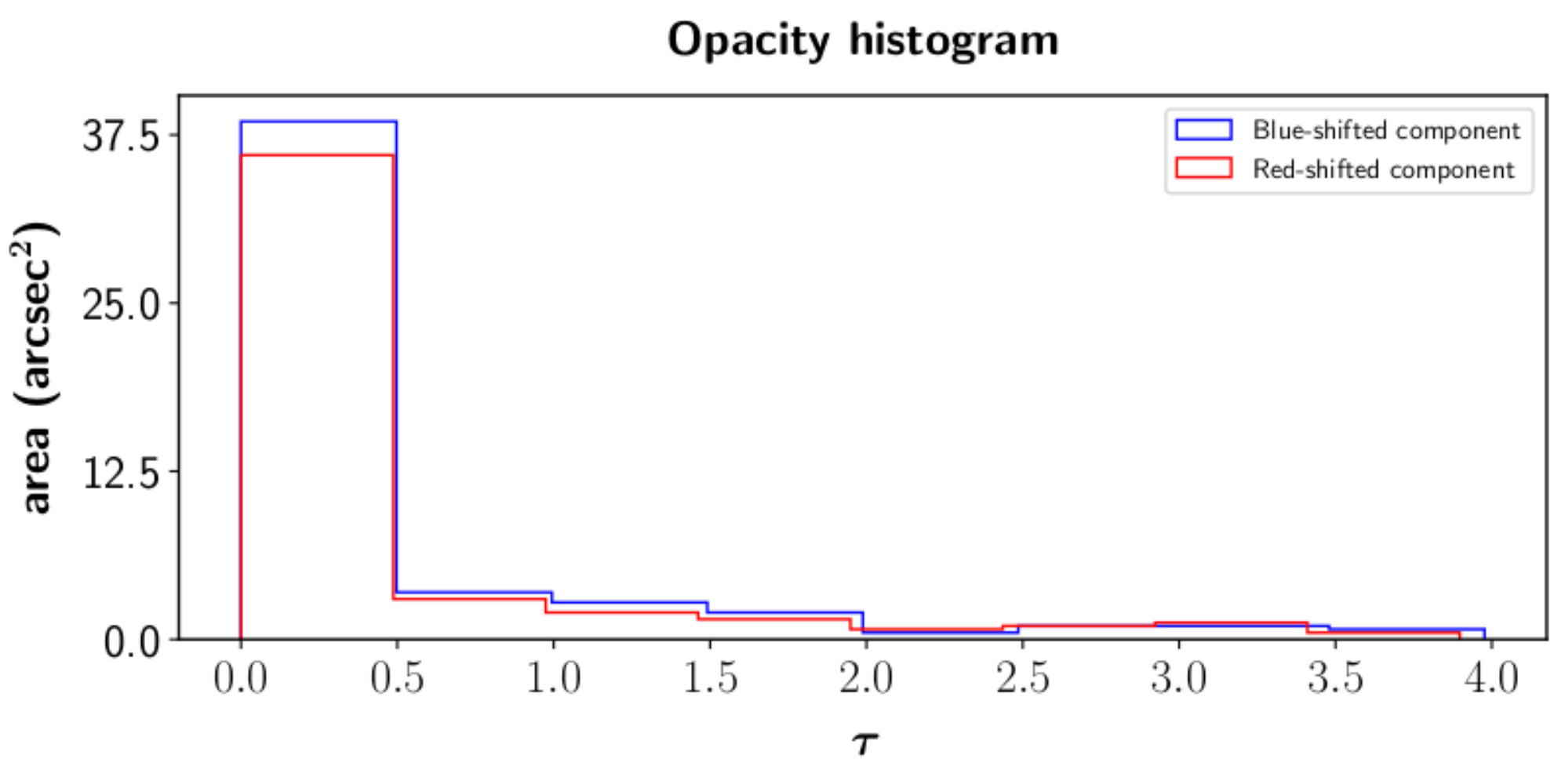} \\
            \end{tabular}
            \caption{Opacity histograms from the fitting of the two individual velocity components.}
            \label{fig:2comp_opacity_hist}
        \end{figure}

        Velocity maps (top images of Fig. \ref{fig:c17o_fullmaps}) show that the two different components, blue and red-shifted, occupy generally two separated regions, at East and West offsets from the center of the source respectively, with some regions overlapping, where the double peak can be observed in the spectra. The mean average, velocity for the two components are 8.1 and 8.6 km~s$^{-1}$ respectively. The velocity dispersion maps (middle images of Fig. \ref{fig:c17o_fullmaps}) show a mean velocity dispersion for the two components of 0.7 and 0.5 km~s$^{-1}$, which get broader, up to 0.9$\sim$1,  closer to the center of the object where the two components overlap (see central spectra in Fig. \ref{fig:lineProf_c17o}). The opacity maps (bottom images of Fig. \ref{fig:c17o_fullmaps}) show that the opacity is generally less than 1, and only goes up to 3 in some specific pixels. These larger values also have a very large error associated, of the order of the value itself, so they are not significant and are not shown in the plots. Figure \ref{fig:2comp_opacity_hist} shows the histograms for the opacity values for both fitted components. The average opacity has been estimated from the \hfs\ modeling and is found to be $\tau_{bs}$ = 0.464 and $\tau_{rs}$ = 0.474, for the blue- and red-shifted components respectively, which is in concordance with the upper limit estimated before. 
        
\section{Discussion} 
    
    \subsection{Linewidths and kinetic temperature}
    
        The main radius probed with the \coseven\ emission was computed from the 3$\sigma$ contour and found to be 860 au. A region enclosing radii from 100 (about half of the FWHM beam) to 860 au has been chosen to analyze the gas kinematics. The kinetic temperature of the gas has been estimated 
        from the formula for dust temperature in an optically thin regime assuming only central heating by the B335 protostar derived by \citet{Shirley2011}. The underlying assumption is that dust and gas are expected to be in thermal equilibrium, being coupled via collisions at the densities probed here ($> 10^{5}$ cm$^{-3}$). We use Eq. \ref{eq:temperature_prof} (\citealt{Evans2015}) which we adapted to the new distance of 164.5 pc.
      
        \begin{equation}
            T_{\rm k}(r) = 67 \, \left[ \frac{r}{\rm 41\ au} \right]^{-0.4} \, {\rm K}
            \label{eq:temperature_prof}
        \end{equation}
        
        The gas kinetic temperature was computed for the two radii probed, with values in range of T$_{k}$(100 au) = 46 K and T$_{k}$(860 au) = 20 K. The observed linewidths obtained in the previous section have been compared with the expected thermal linewidth, given by:
        
        \begin{equation}
            \Delta \nu_{th} = \sqrt{\frac{8\, \ln{2} \, k_{b} \, T_{\rm k}}{m}} \, {\rm km} \, {\rm s^{-1}}
            \label{eq:thermal_linewidth}
        \end{equation}
        
        where T$_{\rm k}$ is the gas kinetic temperature, $m$ is the molecular mass (29.01 amu for \coseven) and $k_{b}$ is the Boltzmann constant. The expected thermal linewidth for \coseven\ has been computed for the temperatures at the two different radii: $\Delta \rm v_{th}$(100 au) = 0.27 km s$^{-1}$ and $\Delta \rm v_{th}$ (860 au) = 0.17 km s$^{-1}$. The observed linewidths are larger than the thermal ones for both velocity components. This indicates that the observed linewidth is the result of the thermal component plus a non-thermal contribution (e.g. turbulence and large-scale motions like infall and outflow, $\rm v_{obs}^{2}$ = $\rm v_{th}^{2}$ + $\rm v_{non-th}^{2}$). The non-thermal contribution of the line has been computed for both velocity components and the results are shown in Table \ref{tab:c17o_kinetics}. The non-thermal component at the inner and outer radius are indistinguishable because of the limited spectral resolution (0.2 km s$^{-1}$).

        The sound speed, $c_{s}$, is in the 0.2--0.3~km~s$^{-1}$ range for temperatures between 20 and 46~ K and $\gamma \sim$ 7/5. This means that the non-thermal contribution to the linewidth is supersonic.
        
        Simulations have shown that in star forming cores systematic large-scale motions (such as infall) can contribute significantly ($\sim$50\%) to the non-thermal component of the linewidth \citep{Guerrero20}. We can do a rough estimation of the contribution from infall, by measuring the infall velocity difference from two different radii, $\Delta\sigma = v_{ff}(r1)-v_{ff}(r2)$, where $v_{ff} (r)= \sqrt{\frac{2GM_{B335}(r)}{r}}$, G is the gravitational constant and M$_{B335}$(r) is the mass enclosed at the considered radius $r$. To compute the mass we consider that the total mass at different radii is the sum of the mass in the envelope plus the mass of the central object. The mass of the gas in the envelope contained up to a certain radius is computed by integrating the 110 GHz dust continuum emission, and using standard assumptions (see Eq. 4 in \citealt{Jorgensen2007}). The maximum velocity difference along the line-of-sight would occur at the core's center. For the two adopted radius, 860 and 100 au, the envelope mass, M$_{env}$, is  $\sim$0.16 \msun\ and $\sim$0.012 \msun, respectively. We also adopt a mass for the central object between 0.05 and 0.26 \msun, which are the predicted values from infall models (see \citealt{Yen2015b}, \citealt{Evans2015}, Evans et al. in prep.). Therefore, the total mass enclosed within a 100 and 860~au radii are in the 0.06--0.16 \msun\ and  0.27--0.41 \msun\ ranges, respectively. The estimated range of $\Delta \nu_{ff}$ is 0.46 - 1.01 km s$^{-1}$. Given the errors coming from the computation of all the previous parameters, the non-thermal contribution of the linewidth is of the same order as the broadening due to free-fall motions along the line of sight: it is hence possible that the observed velocity pattern is due to infall.
        
        \begin{table}
            \centering
            \caption{\coseven\ (1-0) observed linewidth and non-thermal contributions at different radii.}
            \begin{tabular}{c c c}
                \hline
                \hline
                \vspace{-5pt}
                &\\
                 & blue-shifted FWHM & red-shifted FWHM \\
                 & (km~s$^{-1}$) & (km~s$^{-1}$) \\
                \hline
                & \\
                 Observed total & 0.78 & 0.56 \\
                 Non-thermal (100au)  & 0.73 & 0.48 \\
                 Non-thermal (860au) & 0.76 & 0.52 \\
                & \\
                \hline
            \end{tabular}
            \label{tab:c17o_kinetics}
        \end{table}

    \subsection{Possible origins of the observed gas motions}
    
        Our ALMA observations of the \coseven\ (1-0) emission suggest an optically thin emission at all scales probed by the observations (100-860 au). Overall, the spatial extent of the \coseven\ emission is similar to the one of the dust continuum emission, but \coseven\ is less peaked and decreases more smoothly with decreasing density outwards: this suggests that the gas traced with \coseven\ is not mostly related to the outflow cavity. The \coseven\ emission maximum is not coincident with the dust continuum peak position, which might suggest slight abundance variations of the \coseven\ at high densities close to the protostar. Nevertheless, the prominence of the double-peaked velocity pattern does not correlate with the intensity of the dust continuum emission, proving that those profiles are not due to red-shifted absorption against a strong continuum and/or \coseven\ source. These two velocity components thus trace distinguished gas motions. A simple isotropic inside-out envelope collapse can not easily reproduce the gas motions we observe in the B335 envelope. In this section, we discuss various hypothesis for the physical origin of the gas motions observed.
        
        Despite being isolated, B335 is embedded in an extended molecular gas cloud of density $\sim 10^3$ cm$^{-3}$ (\citealt{Frerking1987}). However, \coseven\ is a rare isotopologue which is mostly confined to a high-density central region and its low abundance at large-scales would prevent observing such a tenuous layer, suggesting there is no missing flux coming from large-scale \coseven\ emission. To confirm that this is the case, we estimated the missing flux from the \coeight\ (1-0) ALMA observations by comparing it with the 45 m  Nobeyama data of the same transition presented in \cite{Saito1999}. Our \coeight\ map was smoothed to match the beam size of the Nobeyama telescope, which at the frequency of this transition (109.782 GHz) is 16". We obtained the spectra on a region of one beam size around the center of B335 and transformed the flux density to brightness temperature using the Rayleigh-Jeans law. We obtained a peak temperature of T$_{\rm MB}$ = 0.79 $\pm$ 0.08 K and an integrated temperature of $\int$T$_{\rm MB} dv$ = 0.53 K km/s. This means that our ALMA observations are recovering around 14 $\%$ of the total flux detected with single-dish data. However, because \coseven\ is expected to be much more compact than \coeight\, we expect the missing flux to be much less for the former. \cite{Frerking1987} presented single-dish data of the \coseven\ (1-0) transition and concluded that all their emission is coming from the center of the source in a region smaller than the beam of their telescope (1'.6 for the \coseven\ (1-0) transition). This extension is much smaller than the one observed for \coeight\ (1-0) detected in both works, which is about 4'. This is consistent with the fact that \coseven\ is much less abundant than \coeight, especially at large scales, and that it is mainly tracing the core and not the envelope. Therefore, we expect the missing flux of \coseven\ in our ALMA data to be much less, and to be recovering at least twice the recovered flux of \coeight, i.e. around 30 $\%$. We also note that while our observation might be missing flux, this should not be enough as to produce the huge dip in our data, and it can not explain the structured velocity pattern we observe in the spectral maps, since the missing flux will be at the systemic velocity and would not be able to completely absorb only one of the two components at different offsets.
        
        A possible cause for observing blueshifted and redshifted gas motions in protostellar envelopes could be organized core rotation. Our observations do not support this hypothesis as they do not show a clear velocity gradient in the equatorial plane where rotation motions would mostly contribute to the observed velocity field. Instead, both redshifted and blueshifted velocities are observed in both the northern and southern regions (see Fig. 4). While rotation motions have only been detected at larger envelope radii in B335 ($>$ 2500 au, \citealt{Saito1999}; \citealt{Yen2011}), we stress that the conclusion regarding the absence of small scale rotation (e.g. \citealt{Yen2010}) should be further investigated using the new insights on gas motions in the envelope that our observations have uncovered.
        
        B335 has a well-studied outflow, with its axis close to the plane of the sky and with a well defined X-shaped biconical shape in \co\ \citep{Bjerkeli2019}. Although some contamination by the gas from the outflow cannot be completely ruled out, we present here arguments supporting the hypothesis that our \coseven\ maps can be used to study the envelope gas kinematics. \coseven\ is a rare isotopologue which is known to trace dense envelope gas and is not expected to be detected in more tenuous outflow cavities. The morphology of \coseven\ emission is very different from the one observed in typical outflow cavities tracers, such as \cth\ \citep{Murillo2018} or \co\ (see bottom image in Fig. \ref{fig:C17O_intensity} and \citealt{Bjerkeli2019}). Moreover, no spectral signature of outflow is observed, such as large wings observed in \co\ (\citealt{Bjerkeli2019}), and the maximum velocity shift from the rest velocity remains quite small ($\pm$ 1 km s$^{-1}$). Therefore, the kinematic pattern observed in our \coseven\ maps cannot be produced by outflow alone, and it does provide a strong evidence of distinguished velocity contributions from the gas in the inner region of the B335 protostellar envelope. 
        
        The \coseven\ velocity maps in Fig. \ref{fig:c17o_fullmaps} show that the largest gas velocities are found $\sim1$\arcsec\ from the center along the two northern outflow cavity walls, tracing gas at reverse velocities with respect to the outflow velocities. Considering the 10\degr\ inclination of the system, the spatial distribution of \coseven\ emission following closely that of the dust and other typical dense gas tracers, and the fact that the linewidths of the two velocity components are in general agreement with the expected linewidths from infall motions, the most likely hypothesis is that these high-velocity ($\pm$ 1 km s$^{-1}$) features trace accreting gas flowing along the outflow cavity walls, onto the central protostar. The peak velocities tentatively increase towards the central protostellar objects, for the features along the eastern outflow cavity walls, but no clear velocity gradient could be resolved in the current observations: additional observations with better spatial resolution may allow to test further this hypothesis. Finally, we note that the strongly redshifted emission at the North-East was already detected in ALMA \coeight\ observations reported by \citealt{Yen2015b} (see Fig. 2 in their work).
        
        Dust continuum emission observed with ALMA at various millimeter and sub-millimeter wavelengths all show a striking excess of dust emission associated to the outflow cavity walls. While this could be a temperature effect due to increased heating from the central protostar of these walls, it could also be a true density increase in compact features easily picked up by interferometric observations. 
        Magnetized models of protostellar formation (for a review see \citealt{Zhao2020a}) suggest cavity walls could be preferential sites to develop accretion streamers, as observed in the non-ideal magneto-hydrodynamic (MHD) models of protostellar accretion and outflow launching (\citealt{Machida2014b} or Figure 8 and 9 in \citealt{Tomida2012}). Indeed, these are locations where the poloidal magnetic field is mostly parallel to the inflow direction and therefore would exert less magnetic braking for material infalling along the walls. This hypothesis is also in agreement with the dust polarization observations of magnetic field lines in B335 (the redshifted gas feature we observe along the north-eastern cavity wall is associated to highly organized B-field lines aligned with the tentative gas flow), and the scenario of magnetically-regulated infall proposed in \citet{Maury2018}. 
        
        We note that the observed non-thermal components of the linewidths are found to be supersonic. If the observed gas motions we detect indeed trace localized accretion motions, these could be supersonic. While the development of supersonic filamentary accretion features were reported in numerical models of protostellar formation (\citealt{Padoan2005}; \citealt{Banerjee2006}; \citealt{Kuffmeier2019}), and observations suggested supersonic infall is occurring in a few protostellar envelopes at larger scales ($>1000$ au, \citealt{Tobin2010};  \citealt{Mottram2013}), it is the first time such anisotropic supersonic infall motions are tentatively reported in the B335 inner envelope.

    \subsection{Impact on the characterization of protostellar mass accretion rates}
     
        In the following, we briefly discuss the implications of our work, if the localized accretion features detected in B335 are common while remaining mostly unresolved in many observations of accreting protostars. 
    
        Self-similar solutions for analytical models of the collapse of an isothermal sphere, including  only thermal pressure and gravity, predict typical mass accretion rates of the order $\sim 10^{-4}$ M$_{\odot}$ yr$^{-1}$ (\citealt{Larson1969}; \citealt{Penston1969}; \citealt{Shu1977}). Turbulent models and MHD numerical models have produced slightly lower mass accretion rates $\sim 10^{-6}$--$10^{-5}$ M$_{\odot}$ yr$^{-1}$. Episodic accretion with highly variable rates (from a $\dot{M} \sim$ 10$^{-5}$ M$_{\odot}$ yr$^{-1}$ down to $\dot{M} <$ 10$^{-6}$ M$_{\odot}$ yr$^{-1}$) is often observed in both hydro and MHD numerical models of protostellar formation, in the accretion of envelope material onto the disk and the protostar itself \citep{Lee2021}, and of disk material to the central growing protostar \citep{Dunham2012, Vorobyov2015}. 
        Robust observational estimates of protostellar accretion rates are crucial to distinguish between models, but also to shed light on several open questions on star formation, since they are key quantities for our interpretation of the protostellar luminosities and of the typical duration of the main protostellar accretion phase \citep{Evans2009, Maury2011}. Indeed, observations may have revealed a discrepancy between the observed protostellar bolometric luminosities, and the protostellar accretions rates: this is the so-called `luminosity problem' \citep{Kenyon1990}. Protostellar accretion rates derived from molecular line emission, and more particularly from the modeling of inverse p-Cygni profiles with analytical infall models, should produce luminosities 10-100 times larger than the typically observed bolometric luminosities (for a review, see \citealt{Dunham2014}). Observations of the molecular line profiles in B335 \citep{Evans2005,Evans2015} have been used to fit models of protostellar infall suggesting $\dot{M} \sim 6.2 \times 10^{-6}$ M$_{\odot}$ yr$^{-1}$ (assuming an effective sound speed of 0.3 km/s, Evans et al. in prep.), although arguably these estimates are associated to large error bars. The observed bolometric luminosity of B335 lies an order of magnitude below the accretion luminosity L$_{\rm acc}$ such accretion rates should produce \citep{Evans2015}: despite being a prototype for protostellar infall models, B335 also suffers from the luminosity problem.
     
        If the 'true' protostellar mass accretion rates stems from localized collapsing gas at small scales, potentially affected by unresolved multiple velocity components, the true linewidths associated to infalling gas feeding the growth of the protostar would be quadratically smaller than the ones measured at larger scales where these components would be entangled (or if the individual velocity components are interpreted as being part of a single velocity component with a central dip due to optically thickness). 
        Smaller intrinsic linewidths of the infalling gas, at small radii, may result in smaller effective sound speed and hence lower mass accretion rate derived from the analytical infall models, since $\dot{M} \propto c_{\rm s}^{3}/G$. It is therefore possible that the observed bolometric luminosity of B335 is, ultimately, compatible with its accretion luminosity L$_{\rm acc}$. 
        In the revised B335 scenario we propose here, the mass accretion rate on the protostar could be dictated by localized supersonic infall rather than by the large scale infall rate of the envelope: this may open a window to partially solve the 'luminosity problem', although episodic vigorous accretion would probably still remain necessary to explain the relatively short statistical lifetimes for the Class 0 phase. Future observations should be used to carry out a more detailed characterization of whether a significant fraction of the final stellar mass could be fed to the central object through highly localized anisotropic infall. Recently, \citet{Pineda2020} reported the detection of an 'accretion streamer', connecting the dense core to disk scales, and found a streamer infall rate $\sim 10^{-6}$ M$_{\odot}$ yr-1, of the same order of magnitude as the global mass accretion rate inferred from molecular line observations in B335. Hence, it is possible that many previous studies have failed to grasp the full complexity of the gas motions making up the accretion onto the central protostars. It is the large angular and spectral resolution, along with the great sensitivity of the presented ALMA observations which allowed to detect the two distinct velocity components on the line emission profiles in B335. More observations at the same small-scales and spectral resolution of optically thin emission in different protostars are needed in order to determine if these localized gas motions are common. Moreover, more refined protostellar infall models will have to be carried out in the future, to take this new small-scales into account and include more complex geometries with, e.g., asymmetric structures and preferential accretion along outflow cavities.

\section{Summary}

    ALMA observations of the \coseven\ emission tracing gas kinematics in the B335 envelope have been presented in this work. It is shown that the line emission exhibits widespread double-peaked profiles. From the analysis, the following conclusions have been obtained:
    
        \begin{itemize}
        
            \item Derivations of the line opacity have shown that the emission of the line is optically thin and therefore the observed double-peaked profiles cannot be produced by self-absorption. Therefore, inverse p-Cygni profiles coming from symmetrical inside-out collapse cannot fully explain the observed complex velocity field.
            
            \item After discarding filtering of large-scale emission or other types of motions, such as rotation or outflow, it is determined that only distinct gas motions contributing to the same line-of-sight could explain the observed line profiles pattern. 
            
            \item Linewidth analysis have determined that the two velocity components are compatible with infall motions, and could be due to localized infall in preferential directions. The main hypothesis presented is that the collapse of the envelope onto the protostar is occurring along the equatorial plane but also along the outflow cavity walls, where the magnetic field topology is more favorable.
            
        \end{itemize}
        
    More observations at similar scales and spectral resolution are needed to determine if these double peaked profiles are common in protostellar objects at similar evolutionary states. Moreover, further modeling of the B335 envelope with more complex collapse models, such as anisotropic collapse, are needed to determine which is the exact physical origin of the observed velocity field.

\section*{Acknowledgments}
The authors acknowledge the very useful discussions with N. Evans and Y. Yang.
This project has received funding from the European Research Council (ERC) under the European Union Horizon 2020 research and innovation program (MagneticYSOs project, grant agreement N$\degr$ 679937). J.M.G. is supported by the grant AYA2017-84390-C2-R (AEI/FEDER, UE).
This publication is based on data of ALMA data from the project \AnaelleProj\ (PI. A. Maury).

\vspace{-10pt}
\bibliographystyle{aa}
\bibliography{mybiblio.bib}

\clearpage
\onecolumn 
\begin{appendix}
\section{Technical details of the ALMA observations}
    \begin{table*}[!ht]
    \centering
        \caption{Technical details of the ALMA observations.}
    \begin{tabular}{c c c c c c c c}
        \hline
        &&&&&&& \\
        Config. & Date & Timerange & Center freq. & Spec. Res. & Flux cal. & Phase cal. & Bandpass cal.\\
        & (mm/dd/yy) & (hh:min:s.ms) & (GHz) & (km s$^{-1}$) & &  &   \\
        \hline
        & & & & & & \\
        C40-1 & 19/03/2017 & 10:40:52.8 - 10:53:42.8 & 231.0 & 0.09 & J1751+0939 & J1851+0035 & J1751+0939 \\
        C40-2 & 02/07/2017 & 15:30:38.9 - 15:49:59.2 & 110.0 & 0.08 & J1751+0939 & J1938+0448 & J1751+0939 \\
        C40-4 & 21/11/2016 & 22:41:22.0 - 23:28:19.8 & 231.0 & 0.09 & J2148+0657 & J1851+0035 & J2148+0657 \\
        C40-5 & 10/22/2016 - 10/23/2016 & 23:34:05.4 - 00:02:51.3 & 110.0 & 0.08 & J2148+0657 & J1938+0448 & J2025+3343 \\
        & & & & & & \\
        \hline
    \end{tabular}
    \label{table:technicalInfo}
\end{table*}

\clearpage
\section{\coseven\ full channel maps}

    \begin{figure}[!ht]
        \centering
        \includegraphics[scale=0.7]{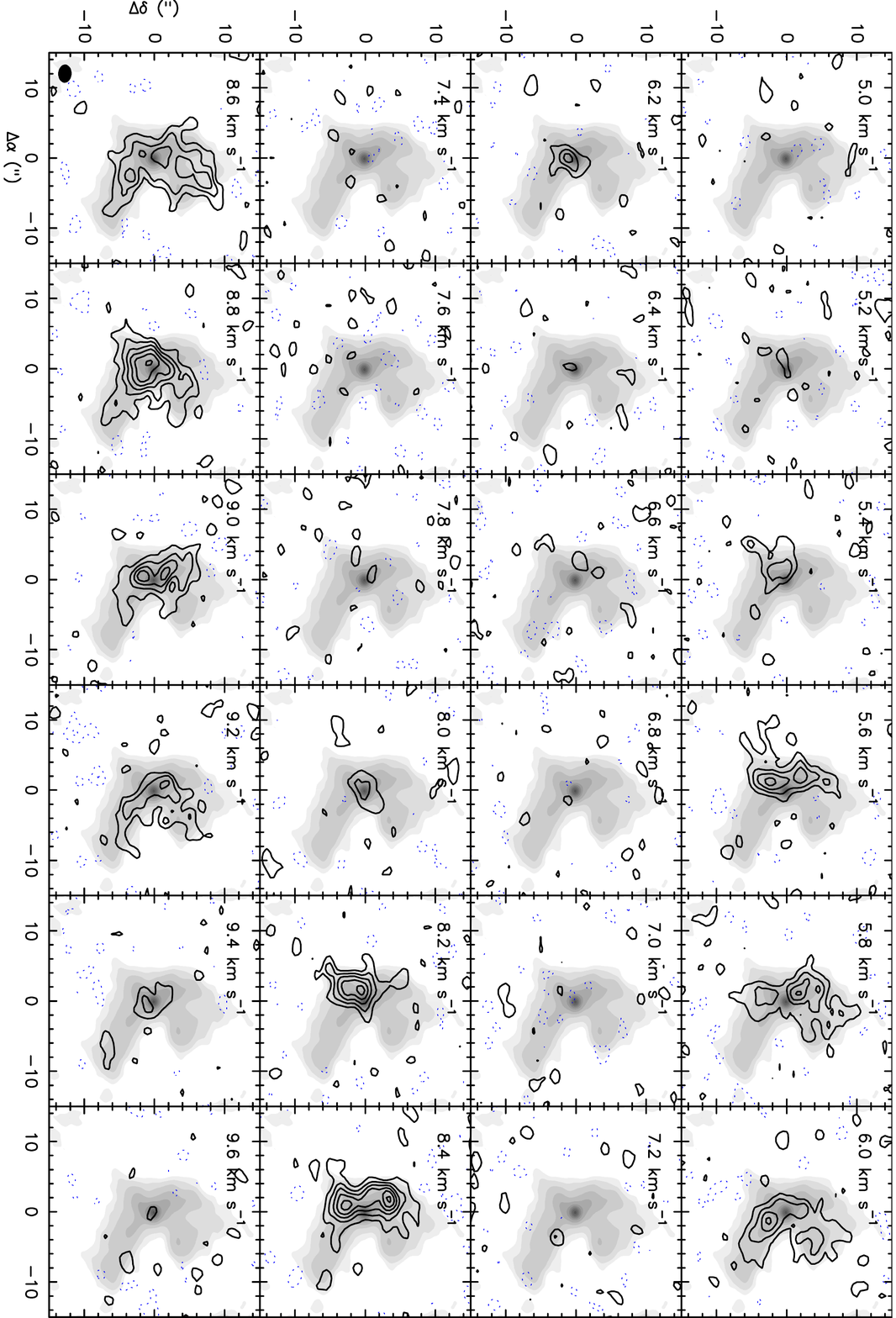}
        \caption{Contour channel maps of the  \coseven\ (1-0) emission showing the two main hyperfine components, overlapped with the gray scale image of the 1.3 mm dust emission. Contours are $-4$, $-2$, and from 2 to 18 by steps of 2 times the rms noise of the channel maps, 10~mJy~Beam$^{-1}$. The synthesized beam is shown in the bottom left panel as a filled ellipse. The $v_{\rm LSR}$ channel velocities are shown in the top left part of the panels.}
        \label{fig:channel_maps_fullHFS}
    \end{figure}

\clearpage
\section{\coeight\ spectral map}
    \begin{figure*}[h!]
        \centering
        \includegraphics[width=\textwidth]{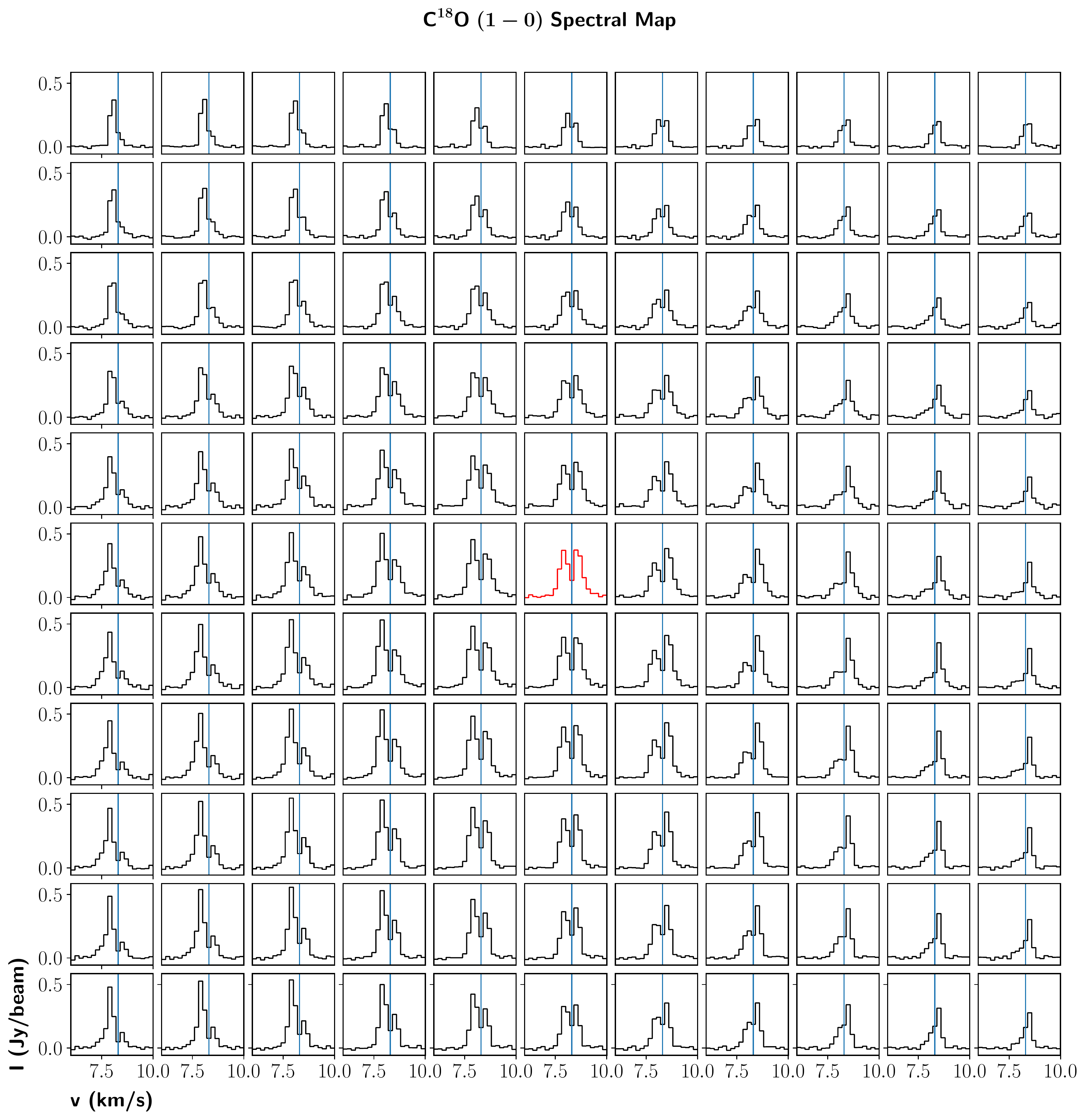} \\
        \caption{Spectral map of the \coeight\ (1-0) emission in the inner 900 au, centered on the dust continuum emission peak. The whole map is 5.5" $\times$ 5.5" and each pixel correspond to 0.5" ($\sim$ 82 au). The green spectra identifies the spectrum at the peak of the continuum emission and the blue line indicates the systemic velocity (8.3 km~s$^{-1}$).}
        \label{fig:lineProf_c18o}
    \end{figure*}

\end{appendix}

\end{document}